\documentclass[aps,pra]{revtex4}
\usepackage{bm}
\usepackage{amsmath}
\usepackage{amssymb}
\usepackage{amsfonts}
\usepackage{graphicx}

\newcommand{\cD}{\mathcal{D}}

\begin{document}

\title{Application of the DRA method to the calculation of the four-loop QED-type tadpoles.}

\author{R.N. Lee}\email{R.N.Lee@inp.nsk.su}
\author{I.S. Terekhov}  \email{I.S.Terekhov@inp.nsk.su}
\affiliation{Budker Institute of Nuclear Physics and Novosibirsk State University,\\ 630090 Novosibirsk, Russia}

\date{\today}

\begin{abstract}
We apply the DRA method to the calculation of the four-loop `QED-type' tadpoles. For
arbitrary space-time dimensionality $\cD$ the results have the form of multiple convergent
sums. We use these results to obtain the $\epsilon$-expansion of the integrals around
$\cD=3$ and $\cD=4$.
\end{abstract}


\maketitle

\section{Introduction}

The calculation of the high-order radiative corrections have become necessary in many areas
of physics, from solid state physics to quantum electrodynamics (QED) and quantum
chromodynamics (QCD). The radiative corrections are expressed in terms of the loop
integrals, therefore  it is necessary to be able to calculate them. Several powerful approaches
to this  problem have been developed. One of the most successful  approach is based on the
integration-by-parts (IBP) reduction procedure \cite{Tkachov1981,Chetyrkin1981}. This
method allows one to reduce the problem of calculation of arbitrary loop integral to that of
some finite set of master integrals. The important feature of the IBP reduction is that it  may also help in the calculation of the master integrals. Namely, using the reduction one can obtain the differential \cite{Kotikov1991,Kotikov1991a,Kotikov1991b,Remiddi1997}  and difference \cite{Tarasov1996,Laporta2000} equations for the master integrals. Recently, in
Ref.~\cite{Lee2010}, the method of calculation based on the dimensional recurrence relation
\cite{Tarasov1996}  and analyticity with respect to space-time dimensionality $\cD$ (the DRA
method) was suggested. This method was applied to the calculation of master integrals for
several physical problems \cite{LeeSmSm2010,LeeSmSm2010a,Lee2010a,LeeSmi2010}. In
these papers the DRA method was combined with other methods such as the method of
Mellin-Barnes representation \cite{Smirnov:1999gc} and sector decomposition method,
implemented in FIESTA \cite{SmiSmTe2009}. 

\begin{figure}
\includegraphics[height=10cm]{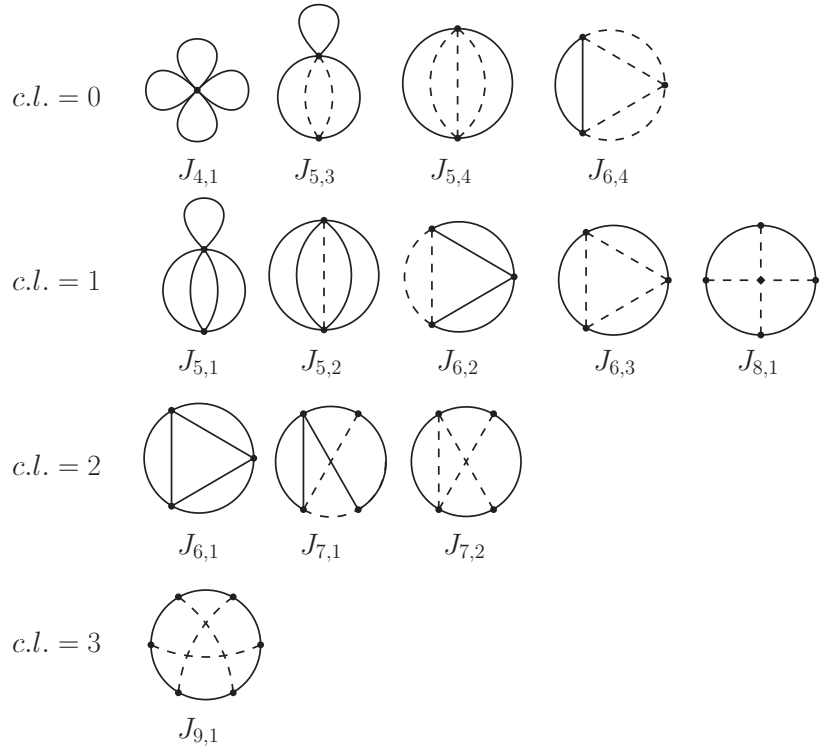}
\caption{Four-loop `QED-type' master integrals considered in this paper. The dashed lines
denote massless propagators $1/k^2$, solid lines denote massive propagators $1/(k^2+1)$.
 }\label{FigMasterIntegrals}
\end{figure}

In the present paper we apply the DRA method
to the calculation of the `QED-type' four-loop master integrals depicted in
Fig.\ref{FigMasterIntegrals}.
These integrals were considered in many papers, see
Refs.~\cite{Broadhurst1996,Laporta2002,Chetyrkin:2004fq,SchrVuo2005,BejdSch2006,KnieKot2006a,KnieKot2006,
KnKoOnV2006,Bejdaki2009,KiriLee2009,Lee2010,Lee2010a}
and references therein. The numerical results for their $\epsilon$-expansions around $\cD=4$
were obtained in Ref.~\cite{SchrVuo2005} using Laporta's difference equation method
\cite{Laporta2000}. Some of the integrals are known in analytic form in terms of the
hypergeometric function. The integrals $J_{6,3}$ and $J_{8,1}$ have been already
investigated using DRA method, see Refs. \cite{KiriLee2009,Lee2010,Lee2010a}. For the
majority of the integrals, several terms of the $\epsilon$-expansion around $\cD=3$ and
$\cD=4$ were found in analytical form in
Refs.~\cite{SchrVuo2005,BejdSch2006,Bejdaki2009}. However, the complete set of the
analytical results for all integrals was not obtained so far. In particular, there is no analytical results for the $\epsilon$-expansion of the most complicated non-planar integral $J_{9,1}$ around $\cD=3$. The goal of the present paper is twofold. First, we demonstrate
some peculiarities of the application of the DRA method appearing in the calculation of the
integrals considered. Second, we present the complete set of the analytical formulas for the
$\epsilon$-expansions of the integrals in Fig.\ref{FigMasterIntegrals} around $\cD=3$ and
$\cD=4$.

\section{The method of calculation}\label{section2}
In order to calculate the integrals depicted in Fig.\ref{FigMasterIntegrals}, we use the DRA
method \cite{Lee2010}, based on the dimensional recurrence relation and analytical
properties of  loop integrals as functions of $\cD$. We evaluate  integrals in the order
determined by their complexity level \cite{LeeSmSm2010}, starting from $c.l.=1$ and ending
with $c.l.=3$. The calculation of the integrals $J_{7,1}(\cD)$ and $J_{7,2}(\cD)$ demands a
slight extension of the approach of Ref.~\cite{Lee2010}, we demonstrate it by presenting here
the calculation of the integral $J_{7,1}(\cD)$. Due to the chosen order of calculation, all
simpler master integrals of $J_{7,1}(\cD)$ (the ones obtained by contracting some lines of
$J_{7,1}(\cD)$) are already known at this stage.

\begin{figure}
\includegraphics[height=1.7cm]{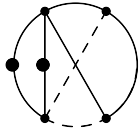}
\caption{The integral $J_{7,1}^a(\cD)$. The dots on the solid
lines denote squared propagators.}\label{J71aFig}
\end{figure}

Obviously, the integral $J_{7,1}(\cD)$ has no infrared divergences for $\cD>2$. Similarly, it
has no ultraviolet divergences for $\cD<7/2$. Therefore, the integral is a holomorphic function
in the stripe $\{\cD|\,\mathrm{Re}\cD\in(2,7/2)\}$. However, this stripe is too narrow to be
chosen as a basic stripe of the DRA method. As it was pointed out in Ref.~\cite{Lee2010}, in
this case one can pass to a new master integral with dots on the massive lines in order to
improve the ultraviolet behavior. A suitable choice is the integral
 $J_{7,1}^a(\cD)$ depicted in Fig. \ref{J71aFig}.
The integral $J_{7,1}^a(\cD)$ has no infrared divergences for $\cD>2$, and no ultraviolet
divergences for $\cD<9/2$, thus, being a holomorphic function in the stripe
$\{\cD|\,\mathrm{Re}\cD\in(2,9/2)\}$.  Due to the IBP identities, the integral $J_{7,1}(\cD)$ can
be expressed via $J_{7,1}^a(\cD)$ and simpler integrals as
\begin{eqnarray}\label{J71viaJ71a}
J_{7,1}(\cD)&=&G(\cD)+H(\cD)= \frac{8 (2 \cD-9)}{3 (\cD-4)^2 (3
\cD-10)}J_{7,1}^a(\cD)+H(\cD)\,,
\end{eqnarray}
where $H(\cD)$ is the linear combination of the integrals $J_{4,1}(\cD)$, $J_{5,1}(\cD)$,
$J_{5,2}(\cD)$, $J_{5,3}(\cD)$, $J_{6,2}(\cD)$, see Appendix. Therefore, we can apply the
DRA method to the calculation of $J_{7,1}^a(\cD)$, and  then use Eq. (\ref{J71viaJ71a}) to
determine $J_{7,1}(\cD)$. However, we find it convenient to apply the DRA method directly to
the integral $J_{7,1}(\cD)$, using Eq. (\ref{J71viaJ71a}) for the determination of the analytical
properties of this integral in the basic stripe $S=\{\cD|\,\mathrm{Re}\cD\in(2,4]\}$. The
singularities of $J_{7,1}(\cD)$  in $S$ are determined by those of two terms in the right hand
side of Eq. (\ref{J71viaJ71a}). The singularities of the function $H(\cD)$ are located at
$\cD=10/3,\, 7/2,\, 4$ and totally fixed by the explicit form of the simpler master integrals
(already calculated at this stage). The only singularities of the first term are the first- and the
second-order poles at $\cD=10/3$ and $\cD=4$, respectively. The principal parts of the
Laurent series expansion of the first term around these two points are not known. However, by
the proper choice of summing factor $\Sigma(\cD)$ we can achieve that the product
$\Sigma(\cD)G(\cD)$ is holomorphic in $S$, see below.

The dimensional recurrence relation for the integral $J_{7,1}(\cD)$ reads
\begin{eqnarray}
J_{7,1}(\cD+2)&=& -\frac{64 (2 \cD-7) (2 \cD-5)}
{3 (\cD-2)^2 (\cD-1) \cD (3 \cD-8) (3 \cD-4)}J_{7,1}(\cD)
+ R(\cD)\, ,\label{Recurrence_J71}
\end{eqnarray}
where the function $R(\cD)$ is given in the Appendix. Choosing the summing factor
$\Sigma(\cD)$ in the form
\begin{eqnarray}\label{solutionSigma1}
\Sigma({\cD})=\frac{2^{7-2 \cD}
\Gamma \left(\frac{9}{2}-\cD\right) \Gamma \left(\frac{\cD}{2}\right)}{\pi ^{5/2} \Gamma \left(5-\frac{3
\cD}{2}\right) \Gamma (3-\cD)}\,,
\end{eqnarray}
we can rewrite Eq. (\ref{Recurrence_J71}) as
\begin{eqnarray}\label{dimensionRecurrRel for g}
g(\cD+2)=g(\cD)+r(\cD)\, ,
\end{eqnarray}
where $g(\cD)=\Sigma(\cD)J_{7,1}(\cD)$, $r(\cD)=\Sigma(\cD+2)R(\cD)$. It follows from Eq.
(\ref{solutionSigma1}), that the function $\Sigma(\cD)$ has first-order zeros at $\cD=3,\,
10/3$, second-order zero at $\cD=4$, and behaves at $\mathrm{Im} \cD\to \pm \infty$ as
$\exp\{\pi|\cD|/2\}$. Therefore, the product $\Sigma(\cD)G(\cD)$ is a holomorphic function in
$S$ and  falls off exponentially at $\mathrm{Im}\cD\to \pm \infty$. Our choice of $\Sigma(\cD)$
provides that the function $g(\cD)$  has only known singularities in $S$, and falls off
exponentially at $\mathrm{Im} \cD\to \pm \infty$. In order to represent the general solution of
Eq. (\ref{dimensionRecurrRel for g}) in terms of infinite series, see Ref.~\cite{Lee2010}, we
need to decompose the function $r(\cD)$ as
\begin{eqnarray}\label{DecompositionR}
r(\cD)=r_+(\cD)+r_-(\cD+2)\,,
\end{eqnarray}
where $r_\pm(\cD\pm 2k)$ decreases faster then $1/k$ at $k\to \infty$. However, the terms in
$r(\cD)$, proportional to $J_{6,2}(\cD)$ and $J_{5,3}(\cD)$, decrease as $1/k$ at $k\to\infty$,
and the decomposition (\ref{DecompositionR}) is not possible. In order to deal with this
problem  we use the following trick. Let us consider  the dimensional recurrence relation
\begin{eqnarray}\label{dimensionRecurrRel for g tilde}
\tilde{g}(\cD+2)&=&\tilde{g}(\cD)+\tilde{r}(\cD)\,,
\end{eqnarray}
for the linear combination
\begin{equation}\label{definition J}
\tilde{g}(\cD)=\Sigma(\cD)\left(J_{7,1}(\cD)+\alpha(\cD) J_{6,2}(\cD)+\beta(\cD) J_{5,3}(\cD)\right)\,,
\end{equation}
and try to find the rational functions $\alpha(\cD)$, $\beta(\cD)$ such that the terms
proportional to  $J_{6,2}(\cD)$ and $J_{5,3}(\cD)$ in $\tilde{r}(\cD)$ decrease faster than
$1/k$ at large $k$. Using the explicit form of  $\tilde{r}(\cD)$, see Appendix, we choose
$\alpha(\cD)=-1$, and conclude that there is no proper choice of $\beta(\cD)$, and we simply
put $\beta(\cD)=0$. After this, the function $\tilde{r}(\cD)$ can be presented as
\begin{eqnarray}\label{DecompositionRtilde}
\tilde{r}(\cD)&=&\tilde{r}_+(\cD)+\tilde{r}_-(\cD+2)
+\tilde{r}_0(\cD)\,,\\
\tilde{r}_0(\cD)&=&-\frac{1}{\sin ^2\left(\pi  \cD/2\right)}\frac{2}{\cD}\,,
\end{eqnarray}
where $\tilde{r}_\pm(\cD\pm 2k)$ decrease faster then $1/k$ at $k\to \infty$. The function
$\tilde{r}_0(\cD)$  corresponds to the large-$\cD$ asymptotic of the term proportional
$J_{5,3}(\cD)$ in $\tilde{r}(\cD)$. Obviously, $\tilde{r}_0(\cD\pm 2k)\sim 1/k$ at large $k$, so
that the sum
\begin{eqnarray}
\sum_{k=0}^\infty\tilde{r}_0(\cD\pm 2k)
\end{eqnarray}
diverges. Fortunately, the solution of Eq. (\ref{dimensionRecurrRel for g tilde}) with $\tilde{r}$
replaced by $\tilde{r}_0$ can be written explicitly as
\begin{equation}\label{g tildetilde via g53}
\tilde{g}_0(\cD)= -\frac{1}{\sin^2 (\pi \cD/2)}\psi
\left(\frac{\cD}{2}\right)\,,
\end{equation}
where $\psi(x)=\Gamma'(x)/\Gamma(x)$. Therefore, the general solution of the dimensional
recurrence relation (\ref{dimensionRecurrRel for g tilde}) has the form
\begin{eqnarray}\label{g tildetilde solution}
\tilde{g}(\cD)=\omega(z)-\sum_{k=0}^{\infty}\tilde{r}_+(\cD+2k)+\sum_{k=0}^{\infty}\tilde{r}_-(\cD-2k)+\tilde{g}_0(\cD)\,,
\end{eqnarray}
where $\omega(z)=\omega\left(\exp[i\pi\cD]\right)$ is arbitrary periodic function. It follows from
Eq. (\ref{g tildetilde solution}) that the analytical properties of the function $\omega(z)$ are
determined by those of $\tilde{g}$, $\tilde{r}_\pm$, and $\tilde{g}_0$. Namely, $\omega(z)$ is
a meromorphic function, which has poles at $z=\pm i\,,\,\pm 1$, and falls off at $|z|\to\infty$.
Since the principal parts of the Laurent series expansions around these points are determined
by known integrals only, and not by $J_{7,1}$, the function $\omega(z)$ can be easily found
(for brevity, we do not present its explicit form here). Finally, using Eqs.
(\ref{dimensionRecurrRel for g}), (\ref{definition J}), and (\ref{g tildetilde solution}) we obtain
\begin{eqnarray}\label{resultJ71}
J_{7,1}(\cD)=J_{6,2}(\cD)+\Sigma^{-1}(\cD)\left(\omega(\cD)-\sum_{k=0}^{\infty}\tilde{r}_+(\cD+2k)+\sum_{k=0}^{\infty}\tilde{r}_-(\cD-2k)-\frac{1}{\sin ^2(\pi \cD/2)}\psi \left(\frac{\cD}{2}\right)\right)\,.
\end{eqnarray}
This equation is valid for arbitrary $\cD$, and, in particular, can be used for the numerical
calculation of $\epsilon$-expansion of $J_{7,1}(\cD)$ around $\cD=3,\,4$. In  two following
sections we present such expansions for all integrals, depicted in Fig.\ref{FigMasterIntegrals}
The analytical form of the expansions is obtained from the high-precision numerical results
using PSLQ algorithm Ref.~\cite{Bailey}, as implemented in MPFUN multiple-precision
subroutines \cite{Bailey1}.  The coefficients in the expansions are expressed in terms of the
 following transcendental numbers:
 \begin{eqnarray}
\zeta_n=\sum_{k=1}^{\infty}\frac{1}{k^n}\,,\quad a_n=\sum_{k=1}^{\infty}\frac{1}{2^k k^n}=
\mathrm{Li}_n(1/2)\,,\quad
s_6=\sum_{m=1}^\infty\sum_{k=1}^{m}\frac{(-1)^{m+k}}{m^5k}= 0.98744\ldots
 \end{eqnarray}

\section{Expansion around $D=4$}\label{section3}
\subsection*{The integrals with {\it c.l.}$=0$}
\begin{eqnarray}\label{J41_expansion}
\frac{J_{4,1}(4-2\epsilon)}{\Gamma^4(-1+\epsilon)}&=&1\,,
\end{eqnarray}
\begin{eqnarray}\label{J53_expansion}
\frac{J_{5,3}(4-2\epsilon)}{\Gamma^4(-1+\epsilon)}&=&\frac{2^{1-2 \epsilon } \Gamma (2-\epsilon ) \Gamma \left(\epsilon -\frac{1}{2}\right) \Gamma (3 \epsilon -2)}{\Gamma (\epsilon -1) \Gamma \left(2 \epsilon
   -\frac{1}{2}\right)}\,,
\end{eqnarray}
\begin{eqnarray}\label{J54_expansion}
\frac{J_{5,4}(4-2\epsilon)}{\Gamma^4(-1+\epsilon)}&=&-\frac{3\cdot 2^{3-4 \epsilon } (\epsilon -1) \Gamma (2-\epsilon )^2 \Gamma \left(\epsilon +\frac{1}{2}\right) \Gamma (3 \epsilon
   -2) \Gamma (4 \epsilon -3)}{\Gamma (\epsilon )^3 \Gamma \left(3 \epsilon -\frac{1}{2}\right)}\,,
\end{eqnarray}
\begin{eqnarray}\label{J64_expansion}
\frac{J_{6,4}(4-2\epsilon)}{\Gamma^4(-1+\epsilon)}&=&\frac{\Gamma (2-3 \epsilon ) \Gamma (1-\epsilon )^4 \Gamma (\epsilon )^2 \Gamma (3 \epsilon -1)^2 \Gamma (4 \epsilon
-2)}{\Gamma (2-2 \epsilon )^2 \Gamma (2-\epsilon ) \Gamma (\epsilon -1)^4 \Gamma (6 \epsilon -2)}\,.
\end{eqnarray}
\subsection*{The integrals with {\it c.l.}$=1$}
\begin{eqnarray}\label{J51_expansion}
\frac{J_{5,1}(4-2\epsilon)}{\Gamma^4(-1+\epsilon)}&=&
\frac{(1
   -\epsilon )^2}{(1-3 \epsilon ) (2-3 \epsilon ) (1-2 \epsilon )}
\Biggl\{
   -4+\frac{44 \epsilon }{3}-\frac{224 \epsilon ^4 \zeta _3}{3}
   +\epsilon ^5 \biggl(\frac{272 \pi ^4}{45}+\frac{64}{3} \pi ^2 \ln ^2\!2\,-\frac{64 \ln ^4\!2\,}{3}-512 a_4\biggr)
   \nonumber\\&&
   -\epsilon ^6 \biggl(\frac{544}{15} \pi ^4 \ln \!2\,+\frac{128}{3} \pi ^2 \ln ^3\!2\,-\frac{128 \ln ^5\!2\,}{5}+3072 a_5-2480 \zeta _5\biggr)
   +\epsilon ^7 \biggl(\frac{64 \pi ^6}{5}+\frac{544}{5} \pi ^4 \ln ^2\!2\,
   \nonumber\\&&+64 \pi ^2 \ln ^4\!2\,-\frac{128 \ln ^6\!2\,}{5}-18432 a_6-7680 s_6+\frac{9760 \zeta _3^2}{3}\biggr)+O(\epsilon ^8)
\Biggr\}\,,
\end{eqnarray}
\begin{eqnarray}\label{J52_expansion}
\frac{J_{5,2}(4-2\epsilon)}{\Gamma^4(-1+\epsilon)}&=&
\frac{(1-\epsilon )^3}{(1-4 \epsilon ) (3-4 \epsilon ) (1-3 \epsilon ) (2-3 \epsilon ) (1-2 \epsilon )}
\Biggl\{
   -6+50 \epsilon -\frac{344 \epsilon ^2}{3}+\frac{3584 \epsilon ^5 \zeta _3}{3}-\epsilon ^6 \biggl(\frac{8704 \pi ^4}{45}
   \nonumber\\&&
   +\frac{2048}{3} \pi ^2 \ln ^2\!2\,-\frac{2048 \ln ^4\!2\,}{3}-16384 a_4\biggr)
   +\epsilon ^7 \biggl(\frac{34816}{15} \pi ^4 \ln \!2\,+\frac{8192}{3} \pi ^2 \ln ^3\!2\,-\frac{8192 \ln ^5\!2\,}{5}
   \nonumber\\&&
   +196608 a_5-174592 \zeta _5\biggr)
   -\epsilon ^8 \biggl(\frac{13312 \pi ^6}{9}+\frac{69632}{5} \pi ^4 \ln ^2\!2\,+8192 \pi ^2 \ln ^4\!2\,-\frac{16384 \ln ^6\!2\,}{5}
   \nonumber\\&&-2359296 a_6-1081344 s_6+\frac{1266688 \zeta _3^2}{3}\biggr)+O(\epsilon ^9)
\Biggr\}\,,
\end{eqnarray}
\begin{eqnarray}\label{J62_expansion}
\frac{J_{6,2}(4-2\epsilon)}{\Gamma^4(-1+\epsilon)}&=&
   \frac{2}{3}+\frac{4 \epsilon }{3}+\frac{2 \epsilon ^2}{3}
   -\epsilon ^3 \biggl(\frac{44}{3}-\frac{16 \zeta _3}{3}\biggr)
   -\epsilon ^4 \biggl(116-\frac{200 \zeta _3}{3}+\frac{4 \pi ^4}{15}\biggr)
   -\epsilon ^5 \biggl(\frac{1928}{3}-\frac{1192 \zeta _3}{3}+\frac{326 \pi ^4}{45}
   \nonumber\\&&
   +\frac{64}{3} \pi ^2 \ln ^2\!2\,-\frac{64 \ln ^4\!2\,}{3}-512 a_4-96 \zeta _5\biggr)
   -\epsilon ^6 \biggl(\frac{9328}{3}-\frac{5864 \zeta _3}{3}+\frac{2126 \pi ^4}{45}+\frac{448}{3} \pi ^2 \ln ^2\!2\,
   \nonumber\\&&
   -\frac{448 \ln ^4\!2\,}{3}-3584 a_4-\frac{2416}{45} \pi ^4 \ln \!2\,-\frac{512}{9} \pi ^2 \ln ^3\!2\,
   +\frac{512 \ln ^5\!2\,}{15}-4096 a_5+2784 \zeta _5+\frac{8 \pi ^6}{21}
   \nonumber\\&&
   -\frac{64 \zeta _3^2}{3}\biggr)+O(\epsilon ^7)\,,
\end{eqnarray}
\begin{eqnarray}\label{J63_expansion}
\frac{J_{6,3}(4-2\epsilon)}{\Gamma^4(-1+\epsilon)}&=&
   \frac{1}{4}+\frac{\epsilon }{2}
   -\epsilon ^3 \biggl(8-\frac{13 \zeta _3}{2}\biggr)
   -\epsilon ^4 \biggl(\frac{241}{4}-4 \zeta _3+\frac{5 \pi ^4}{8}\biggr)
   -\epsilon ^5 \biggl(\frac{669}{2}+36 \zeta _3+\frac{\pi ^4}{5}-\frac{693 \zeta _5}{2}\biggr)
   \nonumber\\&&
   -\epsilon ^6 \biggl(1636+289 \zeta _3-\frac{21 \pi ^4}{5}-72 \zeta _5+\frac{44 \pi ^6}{21}-\frac{241 \zeta _3^2}{2}\biggr)+O(\epsilon ^7)\,,
\end{eqnarray}
\begin{eqnarray}\label{J81_expansion}
\frac{J_{8,1}(4-2\epsilon)}{\Gamma^4(-1+\epsilon)}&=&
\frac{(1
   -\epsilon )^3}{1-2 \epsilon }
\Biggl\{
   5 \epsilon ^3 \zeta _5
   -\epsilon ^4 \biggl(\frac{11 \pi ^6}{378}+7 \zeta _3^2\biggr)+O(\epsilon ^5)
\Biggr\}\,.
\end{eqnarray}
\subsection*{The integrals with {\it c.l.}$=2$}
\begin{eqnarray}\label{J61_expansion}
\frac{J_{6,1}(4-2\epsilon)}{\Gamma^4(-1+\epsilon)}&=&
   \frac{3}{2}+\frac{7 \epsilon }{2}+\frac{9 \epsilon ^2}{2}
   -\epsilon ^3 \biggl(\frac{39}{2}+3 \zeta _3\biggr)
   -\epsilon ^4 \biggl(208-109 \zeta _3+\frac{\pi ^4}{20}\biggr)
   -\epsilon ^5 \biggl(1254-855 \zeta _3+\frac{547 \pi ^4}{60}
   \nonumber\\&&
   +32 \pi ^2 \ln ^2\!2\,-32 \ln ^4\!2\,-768 a_4-189 \zeta _5\biggr)
   -\epsilon ^6 \biggl(6336-4851 \zeta _3+\frac{271 \pi ^4}{4}+240 \pi ^2 \ln ^2\!2\,
   \nonumber\\&&
   -240 \ln ^4\!2\,
   -5760 a_4-\frac{272}{5} \pi ^4 \ln \!2\,-64 \pi ^2 \ln ^3\!2\,+\frac{192 \ln ^5\!2\,}{5}-4608 a_5+3531 \zeta _5
   +\frac{17 \pi ^6}{21}+498 \zeta _3^2\biggr)
   \nonumber\\&&
   +O(\epsilon ^7)\,,
\end{eqnarray}
\begin{eqnarray}\label{J71_expansion}
\frac{J_{7,1}(4-2\epsilon)}{\Gamma^4(-1+\epsilon)}&=&
   -\frac{1}{6}-\frac{5 \epsilon }{6}
   -\epsilon ^2 \biggl(\frac{11}{3}+\zeta _ 3\biggr)
   -\epsilon ^3 \biggl(\frac{44}{3}-\frac{2 \zeta _ 3}{3}+\frac{\pi ^4}{60}\biggr)
   -\epsilon ^4 \biggl(\frac{166}{3}-\frac{31 \zeta _ 3}{3}+\frac{\pi ^4}{6}-53 \zeta _ 5\biggr)
   \nonumber\\&&
   -\epsilon ^5 \biggl(\frac{602}{3}-\frac{38 \zeta _ 3}{3}+\frac{85 \pi ^4}{36}+\frac{16}{3} \pi ^2 \ln ^2\!2\,
   -\frac{16 \ln ^4\!2\,}{3}-128 a_ 4-154 \zeta _ 5+\frac{44 \pi ^6}{189}+128 \zeta _ 3^2\biggr)
   \nonumber\\&&
   +O(\epsilon ^6)\,,
\end{eqnarray}
\begin{eqnarray}\label{J72_expansion}
\frac{J_{7,2}(4-2\epsilon)}{\Gamma^4(-1+\epsilon)}&=&
   -\frac{1}{6}-\frac{5 \epsilon }{6}
   -\epsilon ^2 \biggl(\frac{11}{3}+\frac{\zeta _ 3}{2}\biggr)
   -\epsilon ^3 \biggl(\frac{44}{3}-\frac{13 \zeta _ 3}{6}+\frac{\pi^4}{120}\biggr)
   -\epsilon ^4 \biggl(\frac{166}{3}-\frac{29 \zeta _ 3}{6}+\frac{5\pi ^4}{24}-\frac{43 \zeta _ 5}{2}\biggr)
   \nonumber\\&&
   -\epsilon ^5 \biggl(\frac{602}{3}+\frac{197 \zeta _ 3}{6}+\frac{41
\pi ^4}{120}-\frac{231 \zeta _ 5}{2}+\frac{17 \pi ^6}{189}+\frac{105
\zeta _ 3^2}{2}\biggr)+O(\epsilon ^6)\,.
\end{eqnarray}
\subsection*{The integral with {\it c.l.}$=3$}
\begin{eqnarray}\label{J91_expansion}
\frac{J_{9,1}(4-2\epsilon)}{\Gamma^4(-1+\epsilon)}&=&
\frac{1}{7 \epsilon +1}
\Biggl\{
   \epsilon ^4 \biggl(-\frac{53}{15} \pi ^4 \ln \!2\,-\frac{16}{3} \pi ^2 \ln ^3\!2\,+\frac{16 \ln ^5\!2\,}{5}-384 a_5+\frac{873 \zeta _5}{2}\biggr)
   -\epsilon ^5 \biggl(-\frac{7457 \pi ^6}{1890}
   \nonumber\\&&
   -\frac{124}{3} \pi ^4 \ln ^2\!2\,-\frac{80}{3} \pi ^2 \ln ^4\!2\,
   +\frac{32 \ln ^6\!2\,}{3}+7680 a_6+4032 s_6-\frac{2859 \zeta _3^2}{2}\biggr)
   +O(\epsilon ^6)
\Biggr\}\,.
\end{eqnarray}
The above expansions were considered in
Refs.~\cite{Broadhurst1996,Laporta2002,Chetyrkin:2004fq,SchrVuo2005,BejdSch2006,
KnieKot2006a,KnieKot2006,KnKoOnV2006,Bejdaki2009,Lee2010a}. Our result for $J_{6,2}$ is in full agreement with that of Ref.\cite{KnieKot2006a}. The analytical form of the expansion of all integrals, except the most complicated integral $J_{9,1}$, up to the terms with maximal transcendentality weight equal to $5$ was presented in Ref.~\cite{SchrVuo2005}. The numerical form of the expansion of $J_{9,1}$ was calculated
in the same paper, however, the precision of this calculation was not sufficient for the
application of the PSLQ algorithm. The $\epsilon^0$ term of $J_{9,1}$ in analytical form was calculated in Refs. \cite{KnKoOnV2006,KnieKot2006a}. In Ref.~\cite{Bejdaki2009} the $\epsilon$-expansion around $\cD=4$ of some integrals  was presented up to the terms with maximal transcendentality weight equal to $8$. Our exact expressions can be immediately used for the extraction of even higher terms of $\epsilon$-expansion, but we assume that the practical significance of these terms is questionable.

\section{Expansion around $D=3$}\label{section4}
\subsection*{The integrals with {\it c.l.}$=1$}
\begin{eqnarray}\label{J51_expansion_D=3}
\frac{J_{5,1}(3-2\epsilon)}{\Gamma^4(-1/2+\epsilon)}&=&
\frac{(1-2 \epsilon )^2}{1-6 \epsilon }
\Biggl\{
   \frac{1}{\epsilon }-4 \ln \!2\,
   +\epsilon  \biggl(\frac{2 \pi ^2}{3}+4 \ln ^2\!2\,\biggr)
   -\epsilon ^2 \biggl(\frac{4}{3} \pi ^2 \ln \!2\,+\frac{8 \ln ^3\!2\,}{3}+38 \zeta _3\biggr)
   +\epsilon ^3 \biggl(\frac{44 \pi ^4}{45}
   \nonumber\\&&
   -\frac{16}{3} \pi ^2 \ln ^2\!2\,+8 \ln ^4\!2\,+160 a_4+216 \ln \!2\, \zeta _3\biggr)
   -\epsilon ^4 \biggl(\frac{88}{45} \pi ^4 \ln \!2\,-\frac{32}{9} \pi ^2 \ln ^3\!2\,+\frac{16 \ln ^5\!2\,}{5}-320 a_5
   \nonumber\\&&
   +36 \pi ^2 \zeta _3+216 \ln ^2\!2\, \zeta _3+1445 \zeta _5\biggr)
   +\epsilon ^5 \biggl(\frac{167 \pi ^6}{27}+\frac{88}{45} \pi ^4 \ln ^2\!2\,-\frac{16}{9} \pi ^2 \ln ^4\!2\,+\frac{16 \ln ^6\!2\,}{15}+640 a_6
   \nonumber\\&&
   -1568 s_6+72 \pi ^2 \ln \!2\, \zeta _3+144 \ln ^3\!2\, \zeta _3+1614 \zeta _3^2+5928 \ln \!2\, \zeta _5\biggr)+O(\epsilon ^6)
\Biggr\}
\,,
\end{eqnarray}
\begin{eqnarray}\label{J52_expansion_D=3}
\frac{J_{5,2}(3-2\epsilon)}{\Gamma^4(-1/2+\epsilon)}&=&
\frac{(1-2 \epsilon )^3}{(1-6 \epsilon ) (1-4 \epsilon )}
\Biggl\{
   \frac{7}{4 \epsilon }-8 \ln \!2\,
   +\epsilon  \biggl(\frac{8 \pi ^2}{3}+16 \ln ^2\!2\,\biggr)
   -\epsilon ^2 \biggl(\frac{32}{3} \pi ^2 \ln \!2\,+\frac{64 \ln ^3\!2\,}{3}+108 \zeta _3\biggr)
   \nonumber\\&&
   +\epsilon ^3 \biggl(\frac{316 \pi ^4}{45}+16 \pi ^2 \ln ^2\!2\,+\frac{80 \ln ^4\!2\,}{3}+128 a_4+544 \ln \!2\, \zeta _3\biggr)
   -\epsilon ^4 \biggl(\frac{1264}{45} \pi ^4 \ln \!2\,+\frac{64}{3} \pi ^2 \ln ^3\!2\,
   \nonumber\\&&
   +\frac{64 \ln ^5\!2\,}{3}-512 a_5+\frac{544 \pi ^2 \zeta _3}{3}+1088 \ln ^2\!2\, \zeta _3+3212 \zeta _5\biggr)
   +\epsilon ^5 \biggl(\frac{21928 \pi ^6}{945}+\frac{2528}{45} \pi ^4 \ln ^2\!2\,
   \nonumber\\&&
   +\frac{64}{3} \pi ^2 \ln ^4\!2\,+\frac{128 \ln ^6\!2\,}{9}+2048 a_6-256 s_6+\frac{2176}{3} \pi ^2 \ln \!2\, \zeta _3+\frac{4352}{3} \ln ^3\!2\, \zeta _3+3768 \zeta _3^2
   \nonumber\\&&+13344 \ln \!2\, \zeta _5\biggr)+O(\epsilon ^6)
\Biggr\}\,,
\end{eqnarray}
\begin{eqnarray}\label{J62_expansion_D=3}
\frac{J_{6,2}(3-2\epsilon)}{\Gamma^4(-1/2+\epsilon)}&=&
(1-2 \epsilon )^3
\Biggl\{
   \frac{\pi ^2}{32 \epsilon }-\biggl(-\frac{1}{8} \pi ^2 \ln \!2\,+\frac{7 \zeta _3}{4}\biggr)
   +\epsilon  \biggl(\frac{89 \pi ^4}{1440}-\frac{1}{12} \pi ^2 \ln ^2\!2\,+\frac{\ln ^4\!2\,}{3}+8 a_4\biggr)
   -\epsilon ^2 \biggl(-\frac{89}{360} \pi ^4 \ln \!2\,
   \nonumber\\&&
   +\frac{1}{9} \pi ^2 \ln ^3\!2\,-\frac{4 \ln ^5\!2\,}{15}+32 a_5+\frac{127 \pi ^2 \zeta _3}{48}+\frac{403 \zeta _5}{16}\biggr)
   +\epsilon ^3 \biggl(\frac{13159 \pi ^6}{60480}+\frac{1}{20} \pi ^4 \ln ^2\!2\,+\frac{1}{3} \pi ^2 \ln ^4\!2\,+\frac{8 \ln ^6\!2\,}{45}
   \nonumber\\&&
   +\frac{32 \pi ^2 a_4}{3}+128 a_6-52 s_6-\frac{5}{4} \pi ^2 \ln \!2\, \zeta _3+\frac{253 \zeta _3^2}{4}\biggr)+O(\epsilon ^4)
\Biggr\}\,,
\end{eqnarray}
\begin{eqnarray}\label{J63_expansion_D=3}
\frac{J_{6,3}(3-2\epsilon)}{\Gamma^4(-1/2+\epsilon)}&=&
(1-2 \epsilon )^3
\Biggl\{
   \frac{\pi ^2}{32 \epsilon }-\biggl(-\frac{3}{8} \pi ^2 \ln \!2\,+\frac{21 \zeta _3}{8}\biggr)
   +\epsilon  \biggl(-\frac{23 \pi ^4}{160}+\frac{3}{4} \pi ^2 \ln ^2\!2\,+\frac{3 \ln ^4\!2\,}{2}+36 a_4\biggr)
   -\epsilon ^2 \biggl(\frac{69}{40} \pi ^4 \ln \!2\,
   \nonumber\\&&
   -3 \pi ^2 \ln ^3\!2\,-\frac{18 \ln ^5\!2\,}{5}+432 a_5+\frac{29 \pi ^2 \zeta _3}{16}-\frac{4743 \zeta _5}{16}\biggr)
   +\epsilon ^3 \biggl(-\frac{1391 \pi ^6}{448}-\frac{247}{20} \pi ^4 \ln ^2\!2\,+11 \pi ^2 \ln ^4\!2\,
   \nonumber\\&&
   +\frac{36 \ln ^6\!2\,}{5}+48 \pi ^2 a_4+5184 a_6+1836 s_6+\frac{81}{4} \pi ^2 \ln \!2\, \zeta _3-\frac{5655 \zeta _3^2}{8}\biggr)+O(\epsilon ^4)
\Biggr\}
\,,
\end{eqnarray}
\begin{eqnarray}\label{J81_expansion_D=3}
\frac{J_{8,1}(3-2\epsilon)}{\Gamma^4(-1/2+\epsilon)}&=&
\Biggl\{
   \frac{\pi ^2}{96}
   +\epsilon  \biggl(-\frac{\pi ^2}{96}+\frac{11 \zeta _3}{16}\biggr)
   +\epsilon ^2 \biggl(-\frac{27 \pi ^2}{32}+\pi ^2 \ln \!2\,-\frac{51 \zeta _3}{16}+\frac{271 \pi ^4}{2880}\biggr)
   +\epsilon ^3 \biggl(\frac{907 \pi ^2}{96}-15 \pi ^2 \ln \!2\,
   \nonumber\\&&
   -\frac{291 \zeta _3}{16}-\frac{439 \pi ^4}{2880}+2 \pi ^2 \ln ^2\!2\,+\frac{17 \pi ^2 \zeta _3}{6}+25 \zeta _5\biggr)
   +\epsilon ^4 \biggl(-\frac{6817 \pi ^2}{96}+129 \pi ^2 \ln \!2\,+\frac{4817 \zeta _3}{16}
   \nonumber\\&&
   -\frac{2159 \pi ^4}{320}-30 \pi ^2 \ln ^2\!2\,+\frac{14}{3} \pi ^4 \ln \!2\,
   +\frac{8}{3} \pi ^2 \ln ^3\!2\,-\frac{37 \pi ^2 \zeta _3}{3}-183 \zeta _5+\frac{10279 \pi ^6}{12960}
   +\pi ^4 \ln ^2\!2\,
   \nonumber\\&&
   -\pi ^2 \ln ^4\!2\,-24 \pi ^2 a_4-21 \pi ^2 \ln \!2\, \zeta _3-\frac{293 \zeta _3^2}{4}\biggr)+O(\epsilon ^5)
\Biggr\}\,.
\end{eqnarray}
\subsection*{The integrals with {\it c.l.}$=2$}
\begin{eqnarray}\label{J61_expansion_D=3}
\frac{J_{6,1}(3-2\epsilon)}{\Gamma^4(-1/2+\epsilon)}&=&
(1-2 \epsilon )^3
\Biggl\{
   \frac{\pi ^2}{32 \epsilon }-\biggl(-\frac{1}{8} \pi ^2 \ln \!2\,+\frac{21 \zeta _3}{8}\biggr)
   +\epsilon  \biggl(-\frac{5 \pi ^4}{48}-\pi ^2 \ln ^2\!2\,+\frac{5 \ln ^4\!2\,}{4}+30 a_4+\frac{63}{4} \ln \!2\, \zeta _3\biggr)
   \nonumber\\&&
   -\epsilon ^2 \biggl(\frac{5}{12} \pi ^4 \ln \!2\,-\frac{5}{3} \pi ^2 \ln ^3\!2\,+\frac{13 \ln ^5\!2\,}{5}+108 \ln \!2\, a_4+228 a_5+\frac{45 \pi ^2 \zeta _3}{16}+\frac{63}{4} \ln ^2\!2\, \zeta _3-\frac{1023 \zeta _5}{8}\biggr)
   \nonumber\\&&
   +\epsilon ^3 \biggl(-\frac{5113 \pi ^6}{6720}-\frac{19}{12} \pi ^4 \ln ^2\!2\,+\frac{1}{6} \pi ^2 \ln ^4\!2\,+\frac{19 \ln ^6\!2\,}{15}+18 \pi ^2 a_4+108 \ln ^2\!2\, a_4+648 \ln \!2\, a_5+1560 a_6
   \nonumber\\&&
   +336 s_6+\frac{9}{2} \pi ^2 \ln \!2\, \zeta _3+\frac{21}{2} \ln ^3\!2\, \zeta _3-\frac{693 \zeta _3^2}{16}-\frac{279}{2} \ln \!2\, \zeta _5\biggr)+O(\epsilon ^4)
\Biggr\}\,,
\end{eqnarray}
\begin{eqnarray}\label{J71_expansion_D=3}
\frac{J_{7,1}(3-2\epsilon)}{\Gamma^4(-1/2+\epsilon)}&=&
\frac{(1-2 \epsilon )^3}{4 \epsilon +1}
\Biggl\{
   \biggl(\frac{\pi ^2}{24}-\frac{\ln ^2\!2\,}{2}\biggr)
   +\epsilon  \biggl(\frac{3}{4} \pi ^2 \ln \!2\,+\ln ^3\!2\,-4 \zeta _3\biggr)
   +\epsilon ^2 \biggl(-\frac{\pi ^4}{144}-\frac{23}{12} \pi ^2 \ln ^2\!2\,+\frac{\ln ^4\!2\,}{12}+30 a_4
   \nonumber\\&&
   -\frac{21}{4} \ln \!2\, \zeta _3\biggr)
   +\epsilon ^3 \biggl(\frac{361}{180} \pi ^4 \ln \!2\,+\frac{7}{9} \pi ^2 \ln ^3\!2\,+\frac{12 \ln ^5\!2\,}{5}+28 \ln \!2\, a_4-28 a_5-\frac{13 \pi ^2 \zeta _3}{6}+\frac{213}{4} \ln ^2\!2\, \zeta _3
   \nonumber\\&&
   -\frac{2103 \zeta _5}{16}\biggr)
   +\epsilon ^4 \biggl(\frac{9361 \pi ^6}{11340}-\frac{79}{36} \pi ^4 \ln ^2\!2\,+\frac{1}{18} \pi ^2 \ln ^4\!2\,-\frac{133 \ln ^6\!2\,}{45}-2 \pi ^2 a_4-92 \ln ^2\!2\, a_4-184 \ln \!2\, a_5
   \nonumber\\&&
   +24 a_6-278 s_6-\frac{237}{4} \pi ^2 \ln \!2\, \zeta _3-\frac{199}{2} \ln ^3\!2\, \zeta _3+\frac{3077 \zeta _3^2}{16}-\frac{837}{8} \ln \!2\, \zeta _5\biggr)+O(\epsilon ^5)
\Biggr\}
\,,
\end{eqnarray}
\begin{eqnarray}\label{J72_expansion_D=3}
\frac{J_{7,2}(3-2\epsilon)}{\Gamma^4(-1/2+\epsilon)}&=&
(1-2 \epsilon )^3
\Biggl\{
   \frac{3 \zeta _3}{16}
   +\epsilon  \biggl(-\frac{\pi ^4}{12}-\frac{1}{3} \pi ^2 \ln ^2\!2\,+\frac{\ln ^4\!2\,}{3}+8 a_4+7 \ln \!2\, \zeta _3\biggr)
   +\epsilon ^2 \biggl(\frac{8}{9} \pi ^2 \ln ^3\!2\,-\frac{16 \ln ^5\!2\,}{15}
   \nonumber\\&&
   -32 \ln \!2\, a_4-32 a_5-\frac{35 \pi ^2 \zeta _3}{12}-14 \ln ^2\!2\, \zeta _3+\frac{1089 \zeta _5}{16}\biggr)
   +\epsilon ^3 \biggl(-\frac{15191 \pi ^6}{45360}-\frac{1}{9} \pi ^4 \ln ^2\!2\,-\frac{11}{9} \pi ^2 \ln ^4\!2\,
   \nonumber\\&&
   +\frac{16 \ln ^6\!2\,}{9}+\frac{8 \pi ^2 a_4}{3}+64 \ln ^2\!2\, a_4+128 \ln \!2\, a_5+128 a_6-84 s_6+\frac{7}{3} \pi ^2 \ln \!2\, \zeta _3+\frac{56}{3} \ln ^3\!2\, \zeta _3+\frac{181 \zeta _3^2}{2}
   \nonumber\\&&
   +\frac{651}{4} \ln \!2\, \zeta _5\biggr)+O(\epsilon ^4)
\Biggr\}\,.
\end{eqnarray}
\subsection*{The integral with {\it c.l.}$=3$}
\begin{eqnarray}\label{J91_expansion_D=3}
\frac{J_{9,1}(3-2\epsilon)}{\Gamma^4(-1/2+\epsilon)}&=&
   \biggl(\frac{3}{32}-\frac{3 \ln \!2\,}{128}-\frac{3 \pi ^2}{256}+\frac{9 \ln ^2\!2\,}{64}-\frac{\zeta _ 3}{64}\biggr)
   +\epsilon  \biggl(-\frac{417}{256}+\frac{309 \ln \!2\,}{256}+\frac{37 \pi ^2}{512}-\frac{33 \ln ^2\!2\,}{128}-\frac{9}{64} \pi ^2 \ln \!2\,
   \nonumber\\&&
   -\frac{9 \ln ^3\!2\,}{32}+\frac{9 \zeta _ 3}{8}-\frac{43 \pi ^4}{23040}-\frac{1}{192} \pi ^2 \ln ^2\!2\,+\frac{\ln ^4\!2\,}{192}+\frac{a_ 4}{8}+\frac{7}{64} \ln \!2\, \zeta _ 3\biggr)
   +\epsilon ^2 \biggl(\frac{8313}{512}-\frac{6687 \ln \!2\,}{512}
   \nonumber\\&&
   -\frac{223 \pi ^2}{1024}-\frac{1653 \ln ^2\!2\,}{256}-\frac{29}{128} \pi ^2 \ln \!2\,+\frac{35 \ln ^3\!2\,}{64}-\frac{77 \zeta _ 3}{256}-\frac{517 \pi ^4}{5120}+\frac{21}{128} \pi ^2 \ln ^2\!2\,+\frac{63 \ln ^4\!2\,}{128}
   \nonumber\\&&
   +\frac{63 a_ 4}{16}+\frac{1071}{128} \ln \!2\, \zeta _ 3+\frac{1}{144} \pi ^2 \ln ^3\!2\,-\frac{\ln ^5\!2\,}{120}-\frac{1}{4} \ln \!2\, a_ 4-\frac{a_ 5}{4}-\frac{11 \pi ^2 \zeta _ 3}{128}-\frac{7}{64} \ln ^2\!2\, \zeta _ 3+\frac{251 \zeta _ 5}{256}\biggr)
   \nonumber\\&&
   -\epsilon ^3 \biggl(\frac{131151}{1024}-\frac{80349 \ln \!2\,}{1024}-\frac{2461 \pi ^2}{2048}-\frac{27903 \ln ^2\!2\,}{512}-\frac{2295}{256} \pi ^2 \ln \!2\,-\frac{1447 \ln ^3\!2\,}{128}+\frac{18557 \zeta _ 3}{512}
   \nonumber\\&&
   -\frac{83419 \pi ^4}{92160}-\frac{1435}{768} \pi ^2 \ln ^2\!2\,+\frac{3199 \ln ^4\!2\,}{768}+\frac{2701 a_ 4}{32}+\frac{13249}{256} \ln \!2\, \zeta _ 3+\frac{17}{40} \pi ^4 \ln \!2\,-\frac{29}{32} \pi ^2 \ln ^3\!2\,
   \nonumber\\&&
   +\frac{39 \ln ^5\!2\,}{20}+\frac{405}{8} \ln \!2\, a_ 4+\frac{423 a_ 5}{8}+\frac{699 \pi ^2 \zeta _ 3}{256}+\frac{4311}{128} \ln ^2\!2\, \zeta _ 3-\frac{63189 \zeta _ 5}{512}-\frac{1445 \pi ^6}{96768}-\frac{17}{192} \pi ^4 \ln ^2\!2\,
   \nonumber\\&&
   +\frac{3}{32} \pi ^2 \ln ^4\!2\,-\frac{\ln ^6\!2\,}{144}+\frac{17 \pi ^2 a_ 4}{8}-\frac{1}{4} \ln ^2\!2\, a_ 4-\frac{1}{2} \ln \!2\, a_ 5-\frac{a_ 6}{2}-\frac{23 s_ 6}{8}+\frac{119}{64} \pi ^2 \ln \!2\, \zeta _ 3-\frac{7}{96} \ln ^3\!2\, \zeta _ 3
   \nonumber\\&&
   -\frac{1441 \zeta _ 3^2}{256}+\frac{713}{128} \ln \!2\, \zeta _ 5\biggr)+O(\epsilon ^4)\,.
\end{eqnarray}
Some of the above expansions were considered in
Refs.~\cite{SchrVuo2003,BejdSch2006,Bejdaki2009,Lee2010a}. The $\epsilon$-expansions
for the integrals $J_{5,1}$ and $J_{6,3}$ are in agreement with those obtained in
Refs.~\cite{SchrVuo2003,BejdSch2006} up to the terms considered in these papers. In
Ref.~\cite{Bejdaki2009} some higher terms of the $\epsilon$-expansions of  the integrals
$J_{5,1}$, $J_{5,2}$, $J_{6,2}$, $J_{6,3}$, $J_{7,1}$ were presented. However, we
observed several inconsistencies of the results of Ref.~\cite{Bejdaki2009} with our results. In
particular, the $\epsilon^4$ terms in Eqs.(5.10), (5.22), and (5.28) of Ref.~\cite{Bejdaki2009}
seem to be incorrect.

\section{Conclusion}
In the present paper we apply the DRA method to the calculation of the four-loop `QED-type'
tadpole master integrals. The results obtained are valid for arbitrary $\cD$, and have form of
the convergent multiple sums. For brevity, these results are not presented here, and are
available from the authors upon request. We have presented the $\epsilon$-expansions of the
integrals around $\cD=3,\,4$. The highest transcendentality weight of the expansions (equal to
6), was chosen rather arbitrarily and should be sufficient for physical applications.  Higher
terms of $\epsilon$-expansion for all considered integrals can be easily obtained from our
exact in $\cD$ expressions for the integrals.

\acknowledgments

This work was supported by  Federal special-purpose program "Scientific and
scientific-pedagogical personnel of innovative Russia", RFBR grants Nos.~09-02-00024,
10-02-01238, and DFG grant No. GZ 436 RUS 113/769/0-3. The work of R.L. was also supported through RFBR  grant No.~08-02-01451, I.S. was also supported by the foundation "Dynasty".
The authors gratefully acknowledge the hospitality and financial support during their visit at
Max-Planck-Institute for Quantum Optics, Garching.

\section{Appendix}
\begin{eqnarray}\label{H(D)}
H(\cD)&=& \frac{(\cD-3) \left(5 \cD^3-59 \cD^2+230 \cD-297\right)}
{3 (\cD-4)^3 (3 \cD-10)}J_{6,2}(\cD)+\frac{(\cD-3) (\cD-2) (3 \cD-8) \left(\cD^2-12 \cD+30\right)}{24 (\cD-4)^3 (2 \cD-7) (3
\cD-10)}J_{5,3}(\cD)\nonumber\\
&-&\frac{(2 \cD-5) (3 \cD-11) (3 \cD-8) \left(4 \cD^2-29 \cD+54\right)}{96
(\cD-4)^3 (\cD-3) (3 \cD-10)} J_{5,2}(\cD)\nonumber\\
&+&\frac{(\cD-2) (3 \cD-8) \left(18 \cD^3-215 \cD^2+845 \cD-1098\right)}{96 (\cD-4)^3(\cD-3) (3
\cD-10)} J_{5,1}(\cD)\nonumber\\
&-&\frac{(\cD-2)^3 \left(4 \cD^3+15 \cD^2-229 \cD+450\right)}{192 (\cD-4)^3
(\cD-3)^2(3 \cD-10)} J_{4,1}(\cD)\,.
\end{eqnarray}

\begin{eqnarray}\label{R(D)}
R(\cD)=A^{(6,2)}(\cD)J_{6,2}(\cD)+A^{(5,3)}(\cD)J_{5,3}(\cD)+
A^{(5,2)}(\cD)J_{5,2}(\cD)+A^{(5,1)}(\cD)J_{5,1}(\cD)+
A^{(4,1)}(\cD)J_{4,1}(\cD)\,,
\end{eqnarray}
where the coefficients $A^{(i,j)}(\cD)$ have the form:
\begin{eqnarray}\label{coefficients A}
A^{(6,2)}(\cD)&=& \frac{16 (\cD-3) \left(37 \cD^4-269 \cD^3+689 \cD^2-718 \cD+246\right)}{3
(\cD-2)^2 (\cD-1)^2 \cD (2 \cD-5) (2 \cD-3) (3 \cD-8) (3 \cD-4)} \,
,\nonumber\\
A^{(5,3)}(\cD)&=& \frac{8 \left(567 \cD^7-8370 \cD^6+52445 \cD^5-180639 \cD^4+369021 \cD^3-446696
\cD^2+296400 \cD-83088\right)}{9
(\cD-3) (\cD-2)^2 (\cD-1)^2 \cD (2 \cD-5)
(2 \cD-3) (3 \cD-8) (3 \cD-4)^2}  \,  ,\nonumber\\
A^{(5,2)}(\cD)&=& -\frac{8 \left(216 \cD^4-1792 \cD^3+5515 \cD^2-7479 \cD+3780\right)}
{9 (\cD-3) (\cD-2) (\cD-1)^2 \cD (2 \cD-3)
(3 \cD-8) (3 \cD-4)^2}  \,  ,\nonumber\\
A^{(5,1)}(\cD)&=&\frac{16 \left(72 \cD^4-576 \cD^3+1691 \cD^2-2171 \cD + 1044\right)}
{9 (\cD-3) (\cD-2) (\cD-1)^2 \cD (3 \cD-8)
(3 \cD-4)^2} \,  ,\nonumber\\
A^{(4,1)}(\cD)&=& -\frac{4 \left(9 \cD^6-423 \cD^5+3527 \cD^4-12560 \cD^3+22449 \cD^2-19854 \cD+6912\right)}{9 (\cD-3)^2 (\cD-2) (\cD-1)^2 \cD (2 \cD-3) (3 \cD-8) (3 \cD-4)^2}  \,  .
\end{eqnarray}

\begin{eqnarray}
\tilde{r}(\cD)&=&\Sigma(\cD+2)\Biggl\{\left(A^{(6,2)}(\cD)-\alpha(\cD) A^{(7,1)}(\cD)+\alpha(\cD+2) B^{(6,2)}(\cD)\right)J_{6,2}(\cD)\nonumber\\
&&+\left(A^{(5,3)}(\cD)+\alpha(\cD+2) B^{(5,3)}(\cD)-\beta(\cD)A^{(7,1)}(\cD)+\beta(\cD+2)C^{(5,3)}(\cD)\right)J_{5,3}(\cD)\nonumber\\
&&+A^{(5,2)}(\cD)J_{5,2}(\cD)+A^{(5,1)}(\cD)J_{5,1}(\cD)
+ A^{(4,1)}(\cD)J_{4,1}(\cD)\Biggr\}\, ,\label{Rtilde}
\end{eqnarray}
where
\begin{eqnarray}\label{coefficients B}
A^{(7,1)}(\cD)&=&-\frac{64 (2 \cD-7) (2 \cD-5)}
{3 (\cD-2)^2 (\cD-1) \cD (3 \cD-8) (3 \cD-4)} \, ,\nonumber\\
B^{(6,2)}(\cD)&=&-\frac{16 (\cD-3) (\cD-2) }{(\cD-1)^3 \cD (2 \cD-5) (2 \cD-3)} \, ,\nonumber\\
B^{(5,3)}(\cD)&=& \frac{8 \left(37 \cD^4-286 \cD^3+811 \cD^2-996 \cD+446\right)}{3 (\cD-2) (\cD-1)^3 \cD (2 \cD-5) (2 \cD-3) (3 \cD-4)}
\,,\nonumber\\
C^{(5,3)}(\cD)&=&-\frac{64 (2 \cD-5) (2 \cD-3)}{3 (\cD-2) (\cD-1) \cD^2
(3 \cD-4) (3 \cD-2)} \,.
\end{eqnarray}


\end{document}